\documentclass[aps,prl,notitlepage,twocolumn,superscriptaddress,longbibliography,nofootinbib]{revtex4-1}
\pdfoutput=1
\usepackage[utf8]{inputenc}
\usepackage{amsmath,amsthm,amssymb,amsfonts}
\usepackage{scalerel}
\usepackage{mathtools}
\usepackage{graphicx}
\usepackage{subfigure}
\usepackage[normalem]{ulem}
\usepackage[colorlinks = true,
            linkcolor = blue,
            urlcolor  = blue,
            citecolor = blue,
            anchorcolor = blue]{hyperref}
\usepackage{mathrsfs}
\usepackage{bbold}
\usepackage{units}
\allowdisplaybreaks
\usepackage{bm}
\usepackage{braket}
\usepackage{makecell}
\usepackage[nameinlink]{cleveref} 
\usepackage{upgreek}
\usepackage{blindtext}
\usepackage{verbatim}
\usepackage{algorithm}
\usepackage[noend]{algpseudocode}
\makeatother

\usepackage[dvipsnames]{xcolor}
\usepackage{bbm}
\usepackage[bb=boondox]{mathalfa}
\usepackage{array}
\usepackage{multirow}
\usepackage{tabularx}
\usepackage{float}
\usepackage{dcolumn}
\usepackage{slashed}
\usepackage{braket}
\usepackage{verbatim}
\usepackage{multirow}
\usepackage{resizegather}
\usepackage{titlesec}
\usepackage[toc,page]{appendix}
\usepackage[export]{adjustbox}
\newcommand{\h}{$\mathcal{H}\,$}
\newcommand{\cavg}{c_{\text{avg}}}
\newcommand{\cvar}{\sigma_{\text{c}}^{2}}

\DeclareMathOperator{\Tr}{Tr}

\begin{document}

\title{{Relaxation Fluctuations of Correlation Functions:\\
Spin and Random Matrix Models}}

\author{Tanay Pathak\,\,\href{https://orcid.org/0000-0003-0419-2583}
{\includegraphics[scale=0.05]{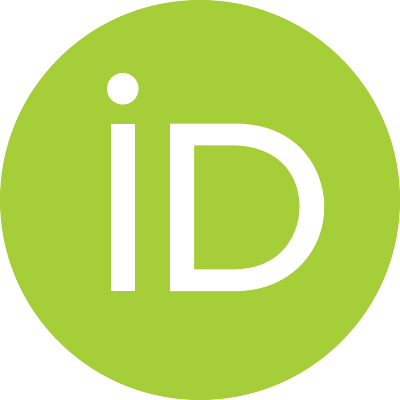}}}
\email{pathak.tanay@yukawa.kyoto-u.ac.jp}
\affiliation{Center for Gravitational Physics and Quantum Information, Yukawa Institute for Theoretical Physics,\\ Kyoto University, Kitashirakawa Oiwakecho, Sakyo-ku, Kyoto 606-8502, Japan}

\begin{abstract}
Spectral statistics and correlations are the usual way to study the presence or absence of quantum chaos in quantum systems. We present our investigation on the study of the fluctuation average and variance of certain correlation functions as a diagnostic measure of quantum chaos and to possibly characterize quantum systems based on it. These quantities are related to eigenvector distribution and eigenvector correlation. Using the Random Matrix Theory certain analytical expressions of these quantities, for the Gaussian orthogonal ensemble case, were calculated before. So as a first step, we study these quantities for the Gaussian unitary ensemble case numerically and obtain certain analytical results for the same. We then carry out our investigations in physical system, such as the mixed-field Ising model. For this model, we find that although the eigenvalue statistics follow the behaviour of corresponding random matrices, the fluctuation average and variance of these correlation functions deviate from the expected random matrix theory behaviour. We then turn our focus on the Rosenzweig-Porter model of the Gaussian Orthogonal Ensemble and Gaussian Unitary Ensemble types. By using the fluctuation average and variance of these correlations, we identify the three distinct phases of these models: the ergodic, the fractal, and the localized phases. We provide an alternative way to study and distinguish the three phases and firmly establish the use of these correlation fluctuations as an alternative way to characterize quantum chaos.
\end{abstract}
\maketitle
~~~~~~~~~~~~Report Number: YITP-24-96


\section{Introduction}
The dynamics of classical systems have been classified into a well-known hierarchy, the ergodic hierarchy \cite{arnold1968ergodic} which ranges from integrable systems to Bernoulli systems. However, the situation is not quite clear in quantum systems \cite{haake1991quantum,stockmann1999quantum} and there is no analogous neat categorization and it is not easy to extend the results of the classical phase space to Hilbert space with ease. The signature of quantum chaos is often attributed to the observation that there is a level repulsion amongst the eigenvalues in quantum Hamiltonians whose classical counterpart is chaotic which is called the Bohigas-Giannoni-Schmit (BGS) conjecture \cite{PhysRevLett.52.1,bohigas1984spectral,casati1980connection,berry1985semiclassical}. This behaviour of eigenvalues and their correlations are well described by the random matrices \cite{mehta2004random,forrester2010log}. On the other hand, we have Berry-Tabor conjecture \cite{berry1977level,marklof2001berry}, for the quantum system whose classical counterpart is integrable and the eigenvalues in quantum systems do not have any correlation, show level clustering and in the semi-classical limit behaves statistically similar to a random process. Similarly eigenstates of chaotic quantum systems are random vectors in the Hilbert space, called the Berry's conjecture \cite{berry1977regular,giannoni1991chaos}. As an example, for a quantum particle, Berry's conjecture states that the eigenstates would be a superposition of plane waves with random phases and Gaussian random amplitude with fixed wavelength. Thus, there is an intricate relationship between random matrices as described using Random Matrix Theory (RMT) and quantum chaos \cite{kriecherbauer2001random}. 
We further note that the resemblance of the energy levels to the RMT hinges on two important features: level repulsion and level rigidity \cite{ullmo2016bohigas}. Level repulsion, characterized by measures such as spacing distribution \cite{mehta2004random} and $\braket{r}$-value \cite{Atas2013distribution}, is an effect observed at short distance, strictly at an energy scale shorter than the mean level spacing $\Delta$ of the system. The corresponding time scale called the Heisenberg time $t_{H}$ is related to the mean energy spacing $\Delta$ of the system, the corresponding energy scale, as $t_{H}= \frac{2 \pi \hbar}{\Delta}$ \cite{haake1991quantum}. On the other hand, the level rigidity is a property of a large energy scale. It is usually characterized by spectral form factor \cite{PhysRevLett.75.4389, Guhr:1997ve,
Leviandier1986, WilkieBrumer1991, Brezin1997, Cotler2017,Gharibyan_2018}, number variance \cite{mehta2004random,haake1991quantum} etc \footnote{Please see \cite{mehta2004random} for a long list of such measures of spectral rigidity.}. The level rigidity is observed at energy scales much larger than the mean level spacings. So it is important to note that while looking for RMT-like behavior it is also crucial to consider the energy scales and the corresponding time scales as well. It is also important to note that even in systems which are chaotic in the classical limit, and have RMT-like levels spacing, deviation from the RMT behaviour has been well observed in their eigenstate statistics \cite{heller1984bound,seligman1984quantum,berry1989quantum,kaplan1998linear,kaplan2000wave,mirlin2000statistics,urbina2007random,bies2002quantum,Smith_2009} associated with various phenomenon. Notable amongst this phenomenon is that of scars where certain unstable periodic orbit in the classical systems scars the quantum eigenstates \cite{heller1984bound} and has also been of interest in both theoretical \cite{Choi_2019, Yao_2022, Turner_2018, Bull_2020, Khemani_2019, Lin_2019, Zhao_2021, Desaules_2022, Langlett_2022, O_Dea_2020, Chandran_2023, Tang_2022, Shibata_2020, Pakrouski_2020, Pakrouski_2021, Dooley_2023, Desaules_2023, Su_2023, Desaules_2023, Magnifico_2020} and experimental studies \cite{Bernien_2017,Zhao_2020,Scherg_2021}.

Over the years, these properties of quantum chaotic systems have been studied in great detail using various tools, not necessarily related to RMT as well. Some of these include (a non-exhaustive list) the eigenstate thermalization hypothesis (ETH) \cite{PhysRevA.43.2046,PhysRevE.50.888,rigol2008thermalization,deutsch2018eigenstate}, entanglement measures \cite{PhysRevLett.119.220603,PhysRevLett.125.180604,Bhattacharjee:2021jff} and measures utilizing operator growth such as out of time-ordered correlator (OTOC) \cite{1969JETPotoc,maldacena2016bound,hashimoto2017out,xu2020does} and recently introduced universal operator growth hypothesis and associated Krylov complexity \cite{Parker:2018yvk, Dymarsky:2019elm, Jian:2020qpp,  Barbon:2019wsy, Rabinovici:2020ryf, Dymarsky:2021bjq,  Caputa:2021sib, Rabinovici:2021qqt, Rabinovici:2021qqt, Balasubramanian:2022tpr, Hornedal:2022pkc, Bhattacharjee:2022vlt,  Bhattacharjee:2022qjw, Erdmenger:2023wjg,Huh:2023jxt, Avdoshkin:2022xuw, Camargo:2022rnt, Bhattacharjee:2024yxj,Baggioli:2024wbz, Balasubramanian:2024ghv}, see \cite{Nandy:2024htc} for a detailed review on recent works. 

With the motivation to study quantum chaotic systems and possibly to classify them, similar to their classical counterpart, we utilize the ideas put forth in \cite{lakshminarayan1994relaxation,PhysRevE.56.2540,lakshminarayan1998relaxation} regarding the relaxation fluctuations of the correlation function and its intrinsic connection to chaos. The study was motivated by and studied using models having well-defined classical limits which are known to be chaotic such as in quantum maps \cite{haake1991quantum,arulnotes,balazs1989quantized,lakshminarayan1994relaxation,PhysRevE.56.2540,lakshminarayan1998relaxation}. By studying the quantum correlation functions which can be related to classical correlations and also to the intricate issues of ergodicity and mixing, an attempt was made to classify quantum dynamics similar to their classical counterpart. Of central interest in this study is the correlation function constructed in such a way that it has a well-defined classical limit. In this paper, we will study these quantities with their original formulation and without necessarily considering the classical limits to illuminate properties of certain quantum spin systems, such as the mixed-field Ising model and other models such as the Gaussian Orthogonal Ensemble (GOE) and Gaussian Unitary Ensemble (GUE) Rosenzweig Porter (RP) models. The mixed field Ising model is a simple model of $1/2-$ spins in a 1D chain with a nearest neighbour interaction and two transverse external fields. The model is known to exhibit both integrable and quantum chaotic limits \cite{PhysRevLett.106.050405,PhysRevLett.111.127205, PhysRevE.91.062128,Roberts:2014isa, PhysRevB.101.174313,Rodriguez-Nieva:2023err,Camargo:2024deu}, depending on the strength of the transverse field. The eigenvalues of the model follow the RMT behaviour, yet the model does not have the expected chaotic dynamics, and exhibit \emph{weak chaos} \cite{PhysRevE.89.012923,PhysRevB.101.174313,PhysRevB.96.060301,PhysRevB.98.144304,PhysRevX.9.031048}. The RP model is a simple one-parameter model of random matrices that has a rich structure and is well studied in literature \cite{PhysRev.120.1698,kravtsov2015random,replicaRP,PhysRevE.108.044211,de2019survival,PhysRevB.109.024205,pino2019ergodic,cadez2024rosenzweig,Bhattacharjee:2024yxj} for its properties. This simple one-parameter model remarkably has a fractal phase (an extended phase) apart from the usually found ergodic (chaotic) and localized (integrable) phases, depending on the value of its parameters. Our findings suggest that even if the eigenvalue statistics follow RMT behaviour it is still possible that the eigenvector correlations do not follow the same, as we will see in the case of the mixed-field Ising model. Such conclusions have also been made for another simple model of quantum chaos, Baker's map \cite{balazs1989quantized,balazs4land,o1992semiclassical,PhysRevE.99.012201}. Furthermore, for models like the RP model, we show that the average and variance of relaxation fluctuations can used to detect the transition quite well.

The outline of the paper is as follows. We start with describing the setup used in Section \ref{sec:setup_rmt} and then discuss the results obtained previously for the GOE random matrices. We then briefly use the method to revisit Baker's map in section \ref{sec:bakerrevisit} and study some of it in some detail. In Section \ref{sec:Ising} we then study the mixed field Ising model within the same framework and describe the similarities of its results with the Baker's map. We then study both the GOE and the GUE-RP model in Section \eqref{sec:rpmodel} and finally end the paper with the conclusion and outlook in Section \ref{sec:conclusion}.

\section{Setup and Random Matrix Theory}\label{sec:setup_rmt}
In this section, we first set up the framework. We will mostly review the procedure already outlined in \cite{lakshminarayan1994relaxation,lakshminarayan1998relaxation,PhysRevE.56.2540}. 

We will be interested in the study of Hilbert space $\mathcal{H}$ of dimension $D$. Let there be two distinct operators on \h, $P_{A}$ and $P_{B}$. For our present purpose, we will focus on the case when these operators are projection operators $P_{A} + P_{B} = \mathbf{I}_{D}$, $\mathbf{I}_{D}$ is the $D$ dimensional identity matrix. The reason for such choice originally was the classical limits of these operators, where classically these correspond to choosing the partition $A$ and $B$ of relevant phase space $\mathcal{P}$ such that $A \cup B= \mathcal{P} $. Consider a unitary evolution operator $U$ which acts on the state in Hilbert space \h of dimension $D$. The quantity of interest considered in \cite{lakshminarayan1994relaxation,lakshminarayan1998relaxation,PhysRevE.56.2540} is the correlation function 
\begin{align}\label{eqn:correlation}
    C_{AB}(t)=  \frac{\Tr(U^{t} P_{A} U^{-t} P_{B})}{D},
\end{align}
which physically captures the overlap of subspace $A$ evolved for time $t$ with the subspace $B$. For classically ergodic systems, when $t \rightarrow \infty$\,, we have $C_{AB} \rightarrow \frac{\Tr(P_{A})\Tr(P_{B})}{D^{2}} = f_{A}f_{B} $ where $f_{A}$ and $f_{B}$ denotes what fraction of total space \h, sub-spaces $A$ and $B$ are respectively. This follows from the fact that for ergodic systems at late times the system becomes uncorrelated, hence the correlation function factorizes. Following \cite{lakshminarayan1998relaxation,PhysRevE.56.2540}, we normalize Eq. \eqref{eqn:correlation}, and hence study the following quantity
\begin{equation}
    c(t) = \frac{C_{AB}(t)}{f_{A}f_{B}}.
\end{equation}
Furthermore, for simplicity, we will choose projectors to be diagonal in the discrete position basis where 
\begin{equation}\label{eq:defPa}
    P_A = \sum_{n}^{f_{A}D} \ket{n} \bra{n},
\end{equation}
and $P_{B} = \mathbf{I}_D - P_{A}$

Since $U$ (or for that matter $H$) has a discrete and finite spectrum, the correlation given by Eq. \eqref{eqn:correlation} is a sum of purely oscillatory terms, hence giving rise to fluctuation in correlation function even after equilibrium has set in. We will mainly be interested in the study of the average and the variance of these fluctuations denoted by $\cavg$ and $\sigma_{c}^{2}$ respectively. We can represent $c(t)$ in terms of eigenfunctions of the evolution operator $U$, denoted by $\psi_{k}$ with the following eigenvalue equation 

\begin{equation}
    U \ket{\psi_{k}} = e^{-i E_{k}} \ket{\psi_{k}}\quad k = 1, \cdots D,
\end{equation}
where $E_{k}$ is the eigenenergy. The correlation function can then be written as 
\begin{align}
    c(t)&= \frac{1}{D f_{A}(1-f_{A})} \sum_{k_{1},k_{2}}\, \sum_{n_{1}=1} ^{D f_{A}}\sum_{n_{2}= D f_{A}+1}^{D} e^{i t (E_{k_2}-E_{k_1})}  \nonumber \\
   & \times \braket{n_{2}|\psi_{k_1}}  \braket{\psi_{k_1}|n_{2}} \braket{n_{1}|\psi_{k_2}}\braket{\psi_{k_2}|n_{2}},
\end{align}
where we have used the fact that $f_{A} +f_{B} = 1$.
The time average of the above quantity, assuming a non-degenerate spectrum, can be simply taken by substituting $k_{1} = k_{2}$. Denoting $f_{A}$ as $f$ we get following after simplification
\begin{align}\label{eqn:cavgfinal}
    \cavg = \frac{1}{D f(1-f)} \sum_{k = 1}^{D}\,\sum_{n_{1}=1} ^{D f}\sum_{n_{2}= D f + 1}^{D} 
    |\braket{n_{1}|\psi_{k}} \braket{n_{2}|\psi_{k}}|^{2}.
\end{align}
Thus, $\cavg$ is a measure of the distribution of eigenvectors in the two sub-spaces $A$ and $B$. Another feature of the average of correlations is that they capture the amount of spread of eigenstates in the subspace $A$ and $B$. If they are equally distributed then the average tends towards the classical time average and is equal to 1. This progression of results towards classically results means the result in the large $D$ limit (notice $\hbar_{\text{eff}}$ is defined to be equal to $\frac{1}{D}$ for such systems and large $D$ implies $\hbar_{\text{eff}} \rightarrow 0$). To capture this feature it is important to look at the scaling of $\cavg$ with $D$.  In \cite{lakshminarayan1994relaxation,lakshminarayan1998relaxation,PhysRevE.56.2540}
it was suggested via various different models that $\cavg< 1$, which implies the reluctance to participate equally in both partitions. However, there is still a possibility of $\cavg = 1$ if there are symmetries present in the systems. One such case where this can happen is, the Baker's map in the presence of parity symmetry. The presence of parity symmetry forces the wave function to be essentially identical in the two subspaces thus giving $c_{avg} = 1$ \cite{PhysRevE.56.2540}. For this reason for all the studies in the paper, we remove all the \emph{unitary} symmetries ( for e.g. parity ) associated with the unitary evolution operator $U$ or Hamiltonian $H$ and always work out the details of $c_{avg}$ and $\sigma_{c}$ using the symmetry resolved blocks.
Similarly the variance of $c(t)$ can be written as follows after simplification
\begin{equation}
    \sigma_{c}^{2} = \frac{2}{D^{2}f^{2}(1-f)^{2}} \sum_{k_{2}>k_{1}} \sum_{n=1}^{f D-1}|\braket{n|\psi_{k_{1}}} \braket{\psi_{k_{2}}|n}|^{4},
\end{equation}
which implies that the variance is a measure of correlations between distinct pairs of eigenvectors. 
\begin{figure}[t]
    \centering
    \includegraphics[width=  \linewidth]{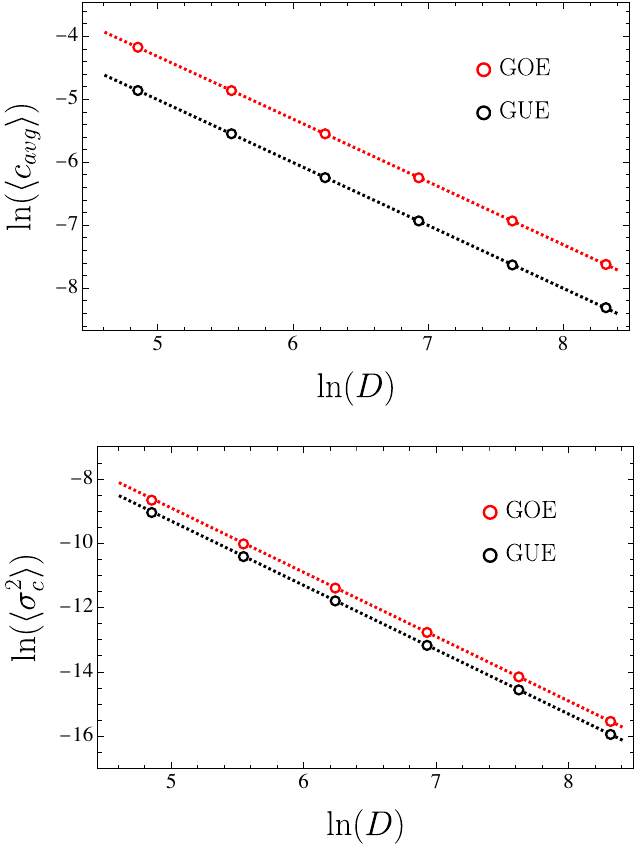}
    \caption{(Top) Deviation of the average of fluctuations from unity with systems size for GOE (red) and GUE (black) matrices. The dotted line corresponds to the analytical result for GOE matrices and the inferred analytic result obtained via best fit for GUE matrices. Shown in the plots are the logarithm of both the deviation as well the system size. (Bottom) The behaviour of the variance of fluctuations with systems size for GOE (red) and GUE (black) matrices. The dashed line corresponds to the analytical result for GOE matrices and the inferred analytic result obtained via best fit for GUE matrices. Shown in the figure is the logarithm of variance as well as system size.}
    \label{fig:rmt_all}
\end{figure}
Using the properties, that the real eigenvectors of GOE matrices are uniformly distributed over a $D-$dimensional sphere and none of the eigenvectors is preferred, it was analytically derived in \cite{lakshminarayan1998relaxation} that the average and variance of the fluctuations are given by 
\begin{align}
    \braket{c_{avg}}_{\text{GOE}} &= \frac{D}{D+2} \sim 1- \frac{2}{D}, \label{eq:cavggoeresult} \\
    \braket{\sigma_{c}^{2}}_{\text{GOE}} &= \frac{3}{D^{2}} + \mathcal{O}\left(\frac{1}{D^{4}}\right), 
\end{align}
where $\braket{\cdot}$ denotes the ensemble average of the quantities.
It is important to note that the above results are independent of the fraction $f$. We have plotted the above results in Fig. \eqref{fig:rmt_all} along with the numerical results obtained using the GOE matrices. We have also plotted the results for the GUE matrices. For numerical purposes for both GOE and GUE type random matrices, we have taken $f= \frac{1}{2}$ and system sizes $D= 128 (2048), 256 (1024), 512 (512), 1024(128), 2048(64)$ $, 4096(32)$, where the numbers in the bracket denote the number of realizations taken for averaging. As outlined in Appendix \ref{appned:guededuce}, we have the following results for GUE matrices
\begin{align}
    \braket{\cavg}_{\text{GUE}} &= \frac{D}{D+1} \sim 1- \frac{1}{D},\\
    \braket{\sigma_{c}^{2}}_{\text{GUE}} &= \frac{2}{D^{2}} + \mathcal{O}\left(\frac{1}{D^{4}}\right).
\end{align}
These are also plotted along with the GUE results in Fig. \eqref{fig:rmt_all} ( the dashed black lines). These behavior of $\braket{\cavg}$ and $\braket{\sigma_{c}^{2}}$ are \emph{universal}, in the sense that they are independent of the partition $f$ used. 
 
\section{Baker's map: A revisit}\label{sec:bakerrevisit}
\begin{figure}[t]
    \centering
    \includegraphics[width=  \linewidth]{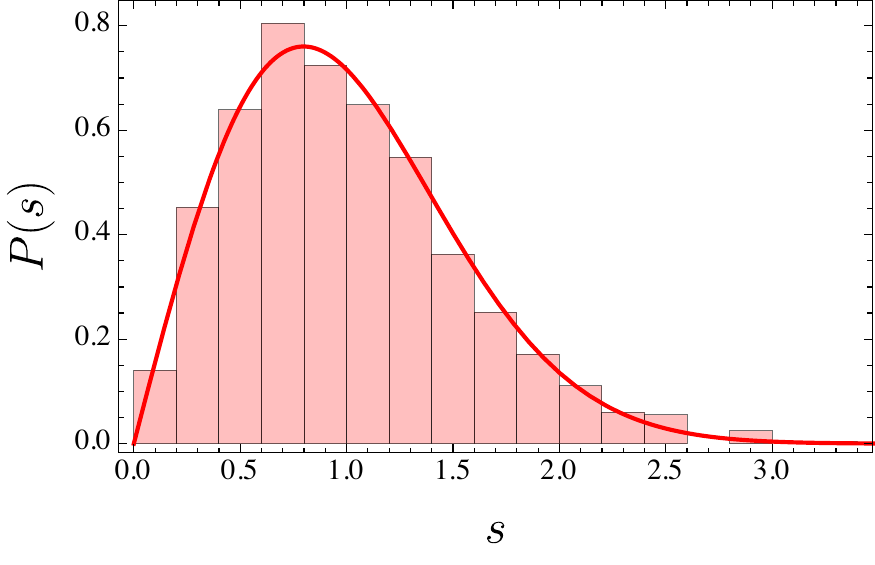}
    \caption{Distribution of eigenangle spacing for Baker's map, with (1/2,1/2) Bernoulli scheme, of size $D = 2000$. The solid line corresponds to the Wigner surmise for GOE. The parity symmetry is resolved and proper unfolding has been done to obtain the histograms.}
    \label{fig:baker_spacing}
\end{figure}

As a brief overview of the setup described in the previous section, we consider a simple quantum model exhibiting chaos, the quantum Baker's map \cite{balazs1989quantized,balazs4land,PhysRevE.56.2540,o1992semiclassical,lakshminarayan1994relaxation,PhysRevE.99.012201,lakshminarayan1998relaxation,arulnotes,haake1991quantum}. Classically the map is defined via the following recursion corresponding to the phase space points $(q_{n},p_{n})$
\begin{equation}
    (q_{n+1}, p_{n+1})= \begin{cases}
    (2 q_{n}, \frac{p_{n}}{2})  \quad \quad \quad \quad~~~ 0< q_{n} < 1/2,\\
    \left(2q_{n}-1, \frac{(p_{n} + 1)}{2} \right)   \quad \frac{1}{2} < q_{n} < 1. \label{eq:bakermap}
  \end{cases}
\end{equation}
The quantized version of the map corresponds to the following unitary operator $U$ \cite{balazs1989quantized}, in position representation, with Hilbert space dimension $D$ 
\begin{equation}\label{eqn:baker_unitary}
    B = G_{D}^{-1} \left(\begin{array}{cc}
G_{D / 2} & 0 \\
0 & G_{D / 2},
\end{array}\right)
\end{equation}
where $G_{D}$ denotes the discrete Fourier transform and is given as 
$(G_{D})_{l,m} = \frac{1}{\sqrt{D}} \exp \left(-2 \pi i (l+\frac{1}{2})(m+\frac{1}{2})\right)$. It is important to note that the Fourier transform is not unique and in general one can have 
\begin{align}\label{eq:fourier}
    (G_{D})_{l,m} = \frac{1}{\sqrt{D}} \exp \left(-2 \pi i (l+\alpha)(m+ \beta)\right)
\end{align}
and different value of $\alpha$ and $\beta$ corresponds to same classical limit with different boundary condition \cite{saraceno1990classical,arulnotes}. It was found in \cite{saraceno1990classical} that the anti-periodic boundary condition $\alpha= \beta= 1/2$, chosen in this paper, maximally preserves the classical reflection symmetry.
The classical map can be shown to be chaotic and mixing with the Lyapunov exponent equal to $\ln(2)$, which is also equal to the Lyapunov exponent obtained using OTOC in the quantized Baker's map \cite{PhysRevE.99.012201}. We remark that we can construct various other kinds of generalized Baker's map similar to Eq. \eqref{eq:bakermap}, by taking into account the different partition schemes in the classical version, e.g. Eq. \eqref{eq:bakermap} is isomorphic to  $(1/2,1/2)$ Bernoulli process \cite{saraceno1990classical}. As another example, we can consider a classical map with Bernoulli scheme $(2/3,1/3)$ \cite{saraceno1990classical,PhysRevE.56.2540,lakshminarayan1998relaxation} given as follows
\begin{equation}
    (q_{n+1}, p_{n+1})= \begin{cases}
    ( \frac{3q_{n}}{2}, \frac{2 p_{n}}{3})  \quad \quad \quad \quad~~~ 0< q_{n} \leq \frac{2}{3},\\
    \left(3q_{n}-2, \frac{(p_{n}+2)}{3} \right)~ \quad \frac{2}{3} < q_{n} < 1. \label{eq:bakermap23}
  \end{cases}
\end{equation}
The corresponding unitary operator is then given by 
\begin{equation}
    B = G_{D}^{-1} \left(\begin{array}{cc}
G_{2 D / 3} & 0 \\
0 & G_{D / 3}.
\end{array}\right)
\end{equation}
The $(1/2,1/2)$ scheme also has a parity symmetry with the parity operator given by $P = -G_{D}^{2}$. The eigenvalues of the unitary operator $U$ are given by $e^{i\theta}$, where $\theta$ is the eigenangle, $\theta \in [0,2 \pi]$ which are uniformly distributed on a circle. The spacing distribution of eigenangles for the chaotic operator follows the GOE distribution. In Fig. \eqref{fig:baker_spacing} we show the eigenangle spacing distribution of Baker's map with Bernoulli scheme $(1/2,1/2)$ after symmetry resolution and unfolding. We observe quite a good agreement with the GOE spacing distribution, which is given by the red solid line in Fig. \eqref{fig:baker_spacing}. We now study the $c_{avg}$ and $\sigma_{c}^{2}$ of the fluctuations for various schemes of Baker's map with $f=1/2$. As an important side note we would like to mention that another way to work without parity symmetry in Baker's map is to choose $\alpha \, \text{or}\, \beta \neq 1/2$ in Eq. \eqref{eq:fourier} thus explicitly breaking the parity symmtery.

\begin{figure}[htbp]
    \centering
    \includegraphics[width=  \linewidth]{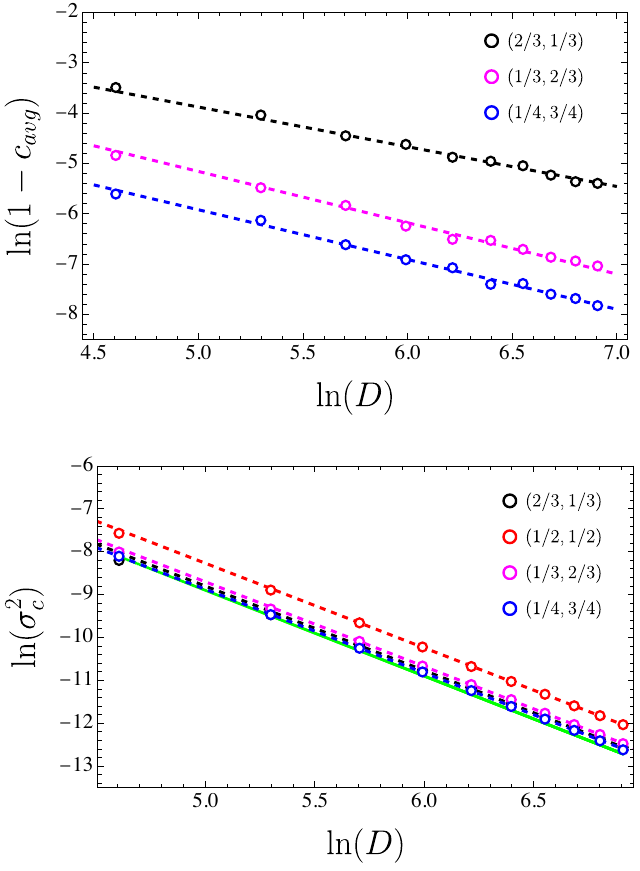}
    \caption{(Top) Logarithmic deviation of $c_{avg}$ from unity for quantum Baker's map for various Bernoulli schemes. Notice that for the $(1/2,1/2)$ scheme, $\cavg =1$ and the corresponding data is not shown. We plot the logarithmic deviation of $\cavg$ from unity. (Bottom) The variance of fluctuation for quantum Baker's amp for various Bernoulli schemes. We show the variation of the logarithmic variance with $\ln(D)$. In both the plots the dashed lines are the best-fit lines obtained using sizes $D > 200$. The green line is the analytical result given by Eq. (9).}
    \label{fig:baker_var_avgl}
\end{figure}

The Fig. \eqref{fig:baker_var_avgl} (Bottom) suggest the following trend of $\cavg$ and $\sigma_{c}^{2}$ with system size $D$
\begin{align}
    \cavg &= 1- \frac{\alpha}{D^{\beta}}, \\
    \sigma_{c}^{2} &= \frac{\alpha_c}{D^{\beta_{c}}}.
\end{align}
For various schemes, we obtain the following values of the best-fit parameter for $\cavg$
\begin{align}
    (\frac{2}{3},\frac{1}{3}) &: \alpha = 1.07 ,\, \beta = .78 \nonumber, \\
    (\frac{1}{4},\frac{3}{4}) &: \alpha = .37 ,\, \beta = .98 \nonumber, \\
    (\frac{1}{3},\frac{2}{3}) &: \alpha = .94 ,\, \beta = 1.02. \nonumber \\
\end{align}

Similarly, for various schemes, we obtain the following values of the best-fit parameter for $\sigma_{c}^{2} $
\begin{align}
    (\frac{2}{3},\frac{1}{3}) &: \alpha_c = 2.84 ,\, \beta_c = 1.97, \nonumber \\
    (\frac{1}{2},\frac{1}{2}) &: \alpha_c = 4.94,\, \beta_c = 1.97, \nonumber \\
    (\frac{1}{4},\frac{3}{4}) &: \alpha_c = 2.39 ,\, \beta_c = 1.95, \nonumber \\
    (\frac{1}{3},\frac{2}{3}) &: \alpha_c = 3.24 ,\, \beta_c = 1.97. \nonumber \\
\end{align}

We observe that for $\sigma_{c}^{2}$ for the Baker's map corresponding to the $(1/2,1/2)$ scheme, the slope is the same as other schemes, $\beta_\sigma \approx -2$, yet it has an offset from other schemes and stands markedly different than other curves which nearly overlap on top of each other. Similar results reporting deviation of $(1/2,1/2)$ scheme from $(2/3,1/3)$ schemes were also reported in \cite{lakshminarayan1998relaxation,PhysRevE.56.2540} and a \emph{possible} reason suggested for such deviation is the presence of a large number of scarred states in Baker's map for this particular scheme \cite{balazs1989quantized}. We merely report these results of Baker's map as an exercise to understand the machinery of the method described. However, it is important to further understand the origin of these results in greater detail.
 
\section{Mixed-Field Ising model}\label{sec:Ising}
We now consider a simple spin chain model that exhibits quantum chaos depending on the values of its parameters, the mixed field Ising model. The model is given as follows 
\begin{align}\label{eqn:hamising}
H = - \sum_{i=1}^{N-1} \sigma_{i}^{z} \sigma_{i+1}^{z} - \sum_{i=1}^{N} ( h_{x}\sigma_{i}^{x}+ h_{z} \sigma_{i}^{z}),
\end{align}
where $\sigma_{i}^{x}$ and $\sigma_{i}^{z}$ are the Pauli matrices at site $i$, defined as 
\begin{align}
    \sigma_{k}^{i}= \mathbf{I}_{2}^{\otimes(i-1)} \otimes \sigma_{k}^{i} \otimes \mathbf{I}_{2}^{\otimes(N-i)}
\end{align}
for $k \in \{x,y,z\}$. We consider open boundary conditions, with $N$ sites with the Hilbert space of dimension $D = 2^{N}$.  For all the numerical computations we choose $f=1/2$.

The model is known to exhibit both chaotic and integrable regimes \cite{PhysRevLett.106.050405,PhysRevLett.111.127205, PhysRevE.91.062128,Roberts:2014isa, PhysRevB.101.174313,Rodriguez-Nieva:2023err,Camargo:2024deu} depending on the strength of longitudinal and transverse fields as follows
\begin{equation}\label{eqn:ising_regime}
   (h_{x},h_{z})=   \begin{cases} 
      (-1.05,0.5), \quad  \text{Chaotic}\\
     (-1,0). \quad \quad \quad  \text{Integrable}
   \end{cases}
\end{equation}
The above chaotic parameters are called the Banuls-Cirac-Hastings (BCH)
parameters \cite{PhysRevLett.106.050405}. We would like to note that there are other parameters in the literature that are claimed to be maximally chaotic based on various other measures of chaos. Some of these parameters are Kim-Huse parameters \cite{PhysRevLett.111.127205} and other parameters suggested in \cite{Rodriguez-Nieva:2023err}. We however choose the former, the BCH parameters and present the result for Kim-Huse parameters in Appendix \ref{appendix:ising}.
Since the model has parity symmetry hence we look for these properties by focusing on symmetry resolves blocks, so as to eradicate the effect of symmetry on these quantities. Once the symmetries are resolved we then carry out our analysis. We also note that for the present case, the projection operators are diagonal in the $\sigma_{z}$ and has been chosen accordingly. This indicates, that all the results, including the RMT predictions in section \ref{sec:setup_rmt}, have a basis dependence, in the sense that we choose a basis such that the projection operators are diagonal.

\begin{figure}[htbp]

   \centering
\includegraphics[width= .95 \linewidth]{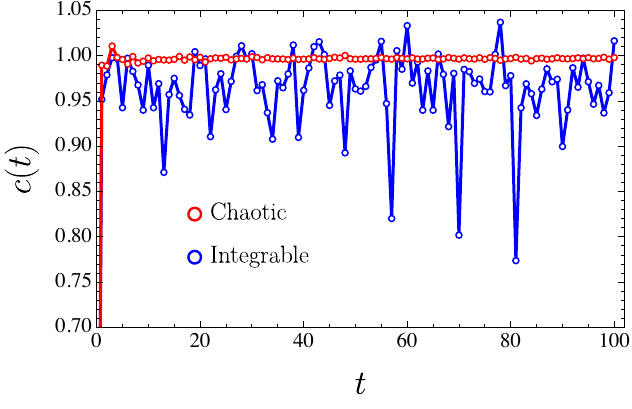}
\caption{Behaviour of $c(t)$ with time for the chaotic and integrable regime of mixed field Ising model for $N= 12$. It is evident from the plot that the fluctuations are larger for the integrable regime than for the chaotic regime.} \label{fig:rf_ising_time}
\end{figure}

In Fig. \eqref{fig:rf_ising_time} we look at the variation of $c(t)$ with time for the two regimes. For practical purposes, and to bypass the need to further resolve the extra symmetry arising for the integrable case, unlike Eq. \eqref{eqn:ising_regime}, we take the parameters to be $(h_{x},h_z) = (-1,0.001)$. We observe that for the integrable regime, the fluctuations are larger than those of the chaotic regime, and there is a clear distinction between the two. 

\begin{figure}[htbp]

   \centering
\includegraphics[width= .95 \linewidth]{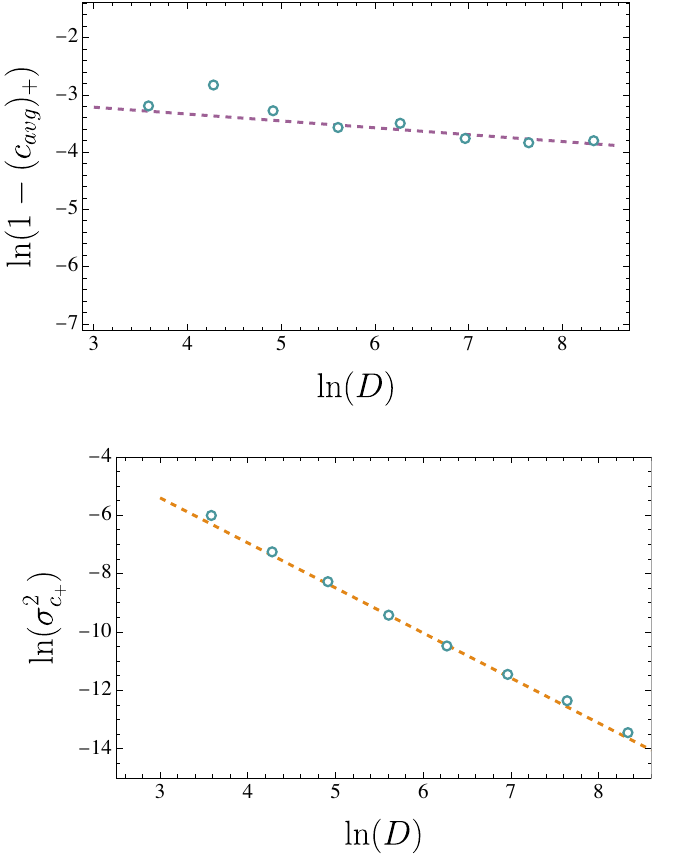}
\caption{(Top) The logarithmic deviation of average fluctuation from unity for the parity +1 sector, with the logarithm of system size, for the mixed field Ising model. We take the Hamiltonian parameters, Eq. \eqref{eqn:hamising}, as $(h_{x},h_{y})= {(-1,.0.001)}$ to avoid extra symmetry arising due to $h_{y}=0$. (Bottom) The behaviour of logarithmic variance with logarithm of system size. Dashed lines show the best-fit line obtained using the data for size $D >200$.} \label{fig:rf_isingint}
\end{figure}

The integrable and chaotic regimes are quite evident from the scaling of $(\cavg)_{+}$ and $\sigma_{+}^{2}$ with $D$, where + sign denotes the +1 parity sector. For the integrable case, the results in Fig. \eqref{fig:rf_isingint}, suggest the scaling behaviour of the following form 
\begin{equation}
(\cavg)_{+} = 1- \frac{\alpha_{+}}{D ^{\beta_{+}}}, 
\end{equation}
with fitting parameters: $ \alpha_{+} = 0.058,\, \beta_{+} = 0.12$. The slope of the best-fit line in Fig (Top).\eqref{fig:rf_isingint} is really small $\sim -0.12$. In fact, we practically observe that for larger $D$ the value saturates implying that as $D$ increases $\cavg$ does not approach unity and is constant. This tendency of $\cavg$ to not scale with $D$ is a \emph{sign of the integrability} of the system. We also explore the variance and observe the following scaling behaviour
\begin{equation}
\sigma^{2}_{+} = \frac{\kappa_{+}}{D ^{\delta_{+}}}, 
\end{equation}
with best-fit parameters $\kappa_{+} = 0.46,\,\delta_{+}= 1.54$. Unlike the behaviour $\cavg$ for the variance, we see that the scaling of variance with $D$. It has been observed in \cite{lakshminarayan1998relaxation,PhysRevE.56.2540} as well that, for integrable systems $\sigma_{c}^{2}$ can scale with $D$ but the coefficient is \emph{non-universal} with a strong dependence on partition $f$. 

\begin{figure}[htbp]

   \centering
\includegraphics[width=   \linewidth]{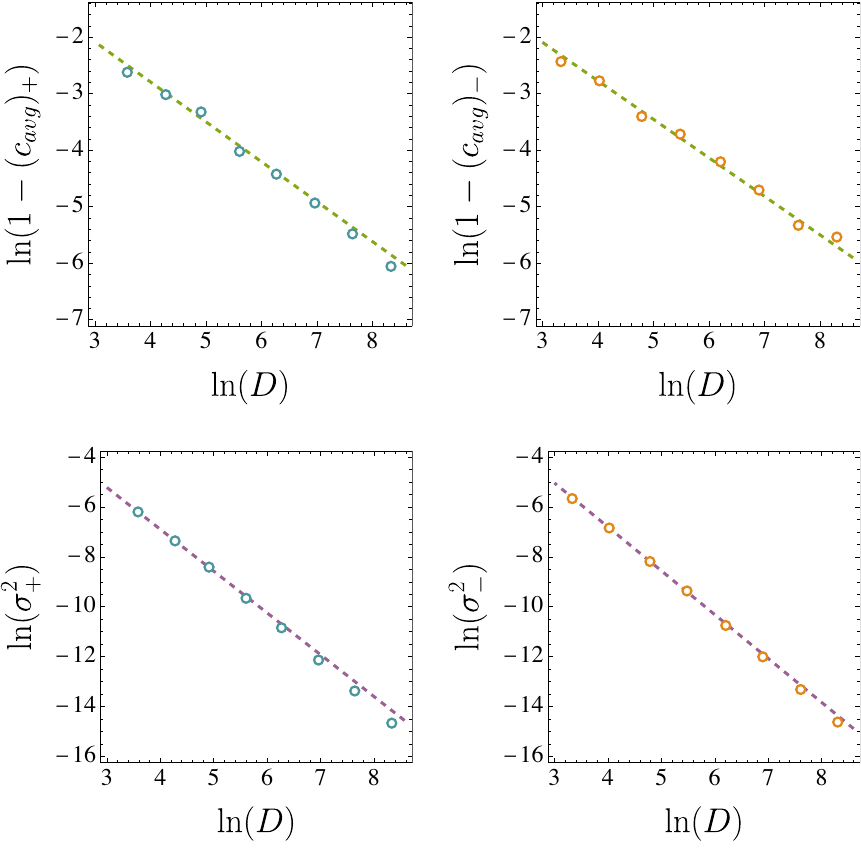}
\caption{(Top) The logarithmic deviation of average fluctuation from unity for the parity $+1$ and $-1$ sector respectively, for the mixed field Ising model. (Bottom) Variation of the logarithmic variance of the relaxation for the mixed field Ising model for parity $+1$ and $-1$ sector respectively. In both the plots the dashed line is the best-fit line. } \label{fig:rf_ising}
\end{figure}

For the chaotic regime, the results for the deviation of average relaxation from unity are shown in Fig (Top). \eqref{fig:rf_ising} for both the parity sectors. We observe the following power-law-like behaviour of average fluctuation with $D$ 
\begin{equation}
(\cavg)_{\pm} = 1- \frac{\alpha_{\pm}}{ D ^{\beta_{\pm}}}, 
\end{equation}
where $\pm$ denotes the $+1$ and $-1$ parity sectors respectively with $\alpha_{\pm} \approx 1,\,\beta_{\pm} \approx \frac{7}{10}$, which deviates from the RMT predictions as discussed in Section \ref{sec:setup_rmt}, where $\alpha= 2$ and $ \beta =1$. This implies that in some sense, the Ising model is less chaotic as compared to other models of quantum chaos which follow the RMT behaviour closely. We have already observed similar features in the case of the quantum Baker's map. These results hint towards a hierarchy among systems exhibiting quantum chaos and suggest that not all systems are as chaotic as others. Furthermore, it shows that eigenvalue statistics are not enough to characterize the dynamics of the quantum systems, as it is possible that although the eigenvalue statistics of these systems follow RMT behaviour closely, the correlation of eigenvectors as discussed in the present context is not in agreement with the corresponding RMT results. We would also like to remark that spin $1/2$ systems are the most \emph{quantum} systems we can have and thus they have peculiar features within the context of quantum chaos as has been reported before in various studies using other measures as well \cite{PhysRevE.89.012923,PhysRevB.101.174313,PhysRevB.96.060301,PhysRevB.98.144304,PhysRevX.9.031048}.

\section{Rosenzweig-Porter Model}\label{sec:rpmodel}
To elucidate the features of integrable to chaotic transition in more detail we now study a random matrix model, the Rosenzweig-Porter (RP) model \cite{PhysRev.120.1698,kravtsov2015random,replicaRP,PhysRevE.108.044211,de2019survival,PhysRevB.109.024205,pino2019ergodic,cadez2024rosenzweig,Bhattacharjee:2024yxj}. The model is given by 
\begin{align}\label{eqn:RPmodel}
    H = A + \frac{1}{D^{\gamma/2}} B\,, 
\end{align}
where $A$ is a diagonal matrix with its elements chosen identically and independently from a standard normal distribution and $B$ is a GOE matrix, whose elements are drawn from a normal distribution with zero means (for both diagonal and off-diagonal elements) and variance $\braket{B_{ii}}^2= 1$ and $\braket{B_{ij}}^2 = 1/2$ for $i \neq j$ respectively. Depending on the value of the free parameter $\gamma$, the model has the following three phases
\begin{align}
    \gamma = \begin{cases}
     [0,1]  \quad ~\mathrm{Ergodic}\,,\\
    [1,2]   \quad~ \mathrm{Fractal}\,,\\
    [2,\infty)~~\, \mathrm{Localized}\,. \label{gamma1}
  \end{cases}
\end{align}

These phases are characterized using the fractal dimensions which are defined using the inverse participation ratio (IPR). For a particular state $\psi_{k}$, we define the eigenstate inverse participation ratio(IPR)  in a basis ${\phi_{i}}$ \cite{edwards1972numerical,misguich2016inverse,tsukerman2017inverse,cadez2023machine} as follows
\begin{equation}\label{eqn:ipr}
    \text{IPR}(\psi_{k}) = \sum_{i}| \braket{\phi_{i}|\psi_{k}}|^{4}.
\end{equation}

The scaling of IPR is given as follows
\begin{align}
    \text{IPR} \sim D^{-f_{d}},
\end{align}
where $f_{d}$ is called the fractal dimensions. For the ergodic phase when the states are uniformly delocalized over all the available states $f_{d}= 1$, however, for the localized phase, it is $f_{d}= 0$ as the states are localized over a single state. The fractal phase is interesting as $f_{d}$ lies between 0 and 1 implying the states are delocalized over the basis, but not fully. 

\begin{figure}[ht]
   \centering
\includegraphics[width=0.85\linewidth]{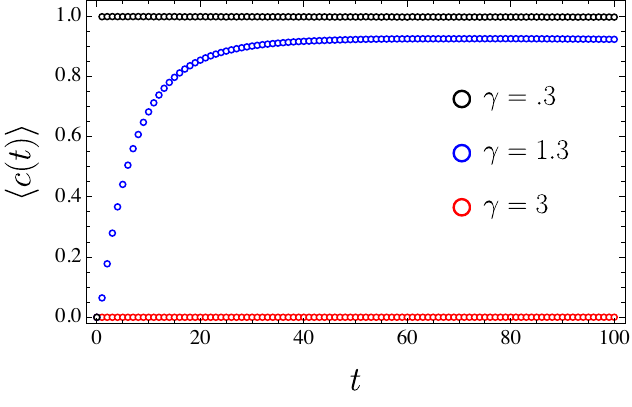}
\caption{The variation of $\braket{c(t)}$ with time $t$ corresponding to three different values of $\gamma$ in the RP model, Eq. \eqref{eqn:RPmodel}. The three values correspond to three different regimes ergodic ( black), fractal (blue) and red (integrable). Results are shown for $D= 512$ with $256$ disordered realizations. } \label{fig:rftimeall3}
\end{figure}
As a side note, it is important to mention that we can also define a basis state IPR, corresponding to a basis state $\phi_{i}$ as 
\begin{equation}
    \text{IPR}(\phi_{i}) = \sum_{k}| \braket{\phi_{i}|\psi_{k}}|^{4}.
\end{equation}
And that the average of basis state IPR over all the basis states is the same as the average of the eigenstate IPR over all the eigenstates.

Here we study these three phases using the properties of relaxation fluctuation, $\braket{c(t)}$, where $\braket{\cdot}$ denotes the ensemble average. In Fig. \eqref{fig:rftimeall3}, we show the behaviour of $\braket{c(t)}$ with time $t$ for three different values of $\gamma= .3, 1.3, 3$ corresponding to ergodic, fractal and integrable regimes respectively. As argued before the effect of chaos is to make the average value of $c(t)$ as close to 1 in the semi-classical limit i.e. large $D$. In the figure we notice that for $\gamma = 0.3$, the ergodic regime the value is close to 1 and for $\gamma= 3$, the integrable limit, it is nearly 0, thus marking the integrable regime. Furthermore, this shows that while the first one is ergodic, the latter is localized. However, we also see that for $\gamma = 1.3$ the value does saturate but at a value intermediate between 1 and 0, which we identify as the fractal regime, where the eigenstates are not fully localized or delocalized. 

\begin{figure}[htbp]
   \centering
\includegraphics[width=0.85\linewidth]{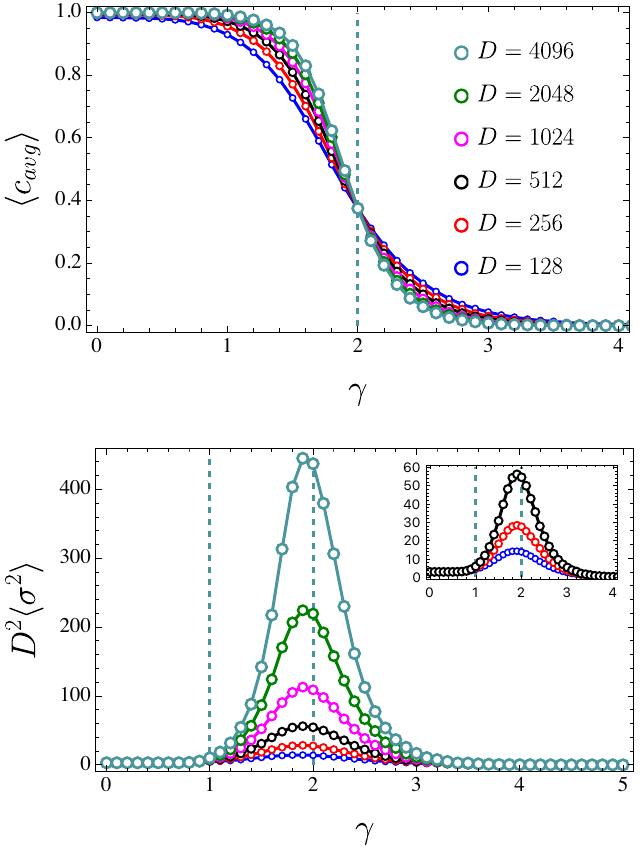}
\caption{(Top) Variation of the average of relaxation fluctuation of GOE RP model. The crossing is visible at $\gamma= 2$,  suggesting a transition at that point. (Bottom) Variation of the re-scaled ( with size) variance of relaxation fluctuation with parameter $\gamma$. The variance shows a clear distinction between the chaotic and the fractal regime. Furthermore, we can identify the localized regime as the one from where the value of rescaled variance starts to drop. Inset shows the zoomed-in view of the first three curves.} \label{fig:rf_all_rpgoe}
\end{figure}

We further study the properties of $\braket{\cavg}$ and $\braket{\cvar}$ for these three regimes with their proper scaling with size $D$ and $f=1/2$, thus giving us the proper insight into the behaviour of these three regimes. In Fig. \eqref{fig:rf_all_rpgoe} (Top) we have numerically studied the variation of the average of the relaxation fluctuation with $\gamma$ for the GOE - RP model. We observe that for small $\gamma$ the model behaves like a GOE-like random matrix and transitions to a localized phase for large $\gamma$ via the fractal phase, where it eventually goes to zero. We have plotted these variations for various system sizes, $D= 128 (1024), 256 (512), 512 (256), 1024 (128), 2048 (64)$ and $4096 (32)$ with the corresponding realizations shown in the brackets, in Fig. \eqref{fig:rf_all_rpgoe}. For the $\braket{\cavg}$ we observe that there is a crossing at $\gamma =2$, capturing the transition from the ergodic to integrable regime. We further observe that the average relaxation fluctuation is incapable of detecting the fractal phase. However, if we study the variance of relaxation fluctuations it is seen there is a clear distinction between the three phases. In Fig. \eqref{fig:rf_all_rpgoe} (Bottom) we have plotted the variation of variation times the system size with $\gamma$. We observe that in both the ergodic (chaotic) as well the localized (integrable) regime the curves collapse. It is also observed that the $D^{2} \braket{\sigma^{2}_{c}}$ is constant till $\gamma \approx 1$, i.e. in the ergodic regime and then gradually increases till $\gamma \approx 2$ in the fractal phase and then further decreases and eventually reaches zero in the deep localized regime. 

\begin{figure}[htbp]
   \centering
\includegraphics[width=0.8\linewidth]{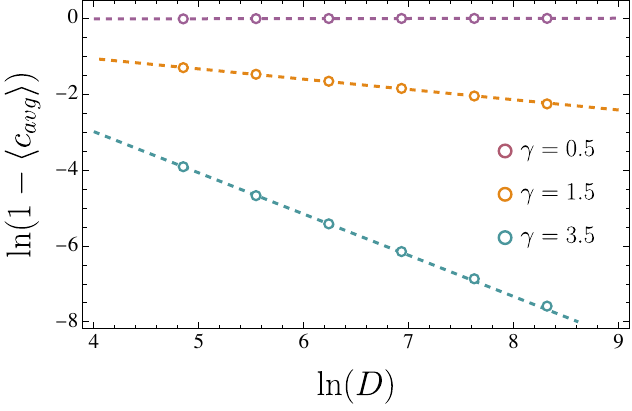}
\caption{ The logarithmic deviation of the average fluctuation from unity for the RP model. We show the variation of deviation with (logarithm of )size. The dashed lines are the best-fit lines. } \label{fig:rf_all_gammarpgoe}
\end{figure}

In Fig. \eqref{fig:rf_all_gammarpgoe} we have plotted the scaling of the logarithm of deviation of the average of relaxation fluctuation from unity with the logarithm of the system size. We choose values $\gamma = 0.5, 1.5, 3.5 $ each corresponding to ergodic, fractal and localized regimes respectively. The plot suggests scaling of $\braket{\cavg}$ of the following form 
\begin{equation}\label{eq:cavgrp}
 \braket{\cavg} = 1- \frac{\alpha}{D^{\beta}},   
\end{equation}
where $\beta = 1.08, 0.27, -0.003$ for the three regimes respectively. It is interesting to note that Eq. \eqref{eq:cavgrp} bears a striking resemblance to the scaling of IPR as defined in Eq. \eqref{eqn:ipr}.

Let us consider the quantity $\cavg$ given by Eq. \eqref{eqn:cavgfinal}
\begin{equation}
     \cavg = \frac{1}{D f(1-f)} \sum_{k = 1}^{D}\,\sum_{n_{1}=1} ^{D f}\sum_{n_{2}= D f + 1}^{D} 
    |\braket{n_{1}|\psi_{k}} \braket{n_{2}|\psi_{k}}|^{2}.
\end{equation}
Consider the case when $f= 1/D$, which implies that partition $A$ is one dimensional. We then have 
\begin{equation}
     \cavg = \frac{1}{(1-1/D)} \sum_{k = 1}^{D}\,\sum_{n_{2}= 2}^{D} 
    |\braket{1|\psi_{k}} \braket{n_{2}|\psi_{k}}|^{2}.
\end{equation}
since we only have one state $\ket{1}$, for subspace $A$.
We rearrange the terms so that we can write 
\begin{align}
  \cavg &= \frac{1}{(1-1/D)} \Big(\sum_{k = 1}^{D}\,\sum_{n_{2}= 1}^{D} 
    |\braket{1|\psi_{k}} \braket{n_{2}|\psi_{k}}|^{2} \nonumber \\
    & -\sum_{k = 1}^{D}\,
    |\braket{1|\psi_{k}}|^{4}\Big)
\end{align}
The second term in the bracket is IPR ( to be precise it is the eigenstate IPR) for a \emph{single basis state} and the quantity $\cavg$ is just a re-scaling of IPR. Thus the quantity $\cavg$ can be viewed as a generalization of IPR when the subspace is spanned by more than one basis states.

For the ergodic ( chaotic) and localized (integrable) these exponents are in accordance with the analytical results from RMT. As we stated earlier in Sec. \ref{sec:setup_rmt}, $\cavg$ captured the amount of spread of the wavefunctions in the two subspaces $A$ and $B$. For a system with chaos, it would mean that this quantity is as close to 1 as possible as $D$ becomes larger and larger. If it is less than 1 then it implies that spread is less and there is \emph{localization} a fact well observed in previous studies as well \cite{lakshminarayan1994relaxation}. To be precise $\beta= 0$ would imply a complete localization and the intermediate values $0 < \beta <1$  would imply a fractal regime.

\begin{figure}[h]
   \centering
\includegraphics[width=0.8\linewidth]{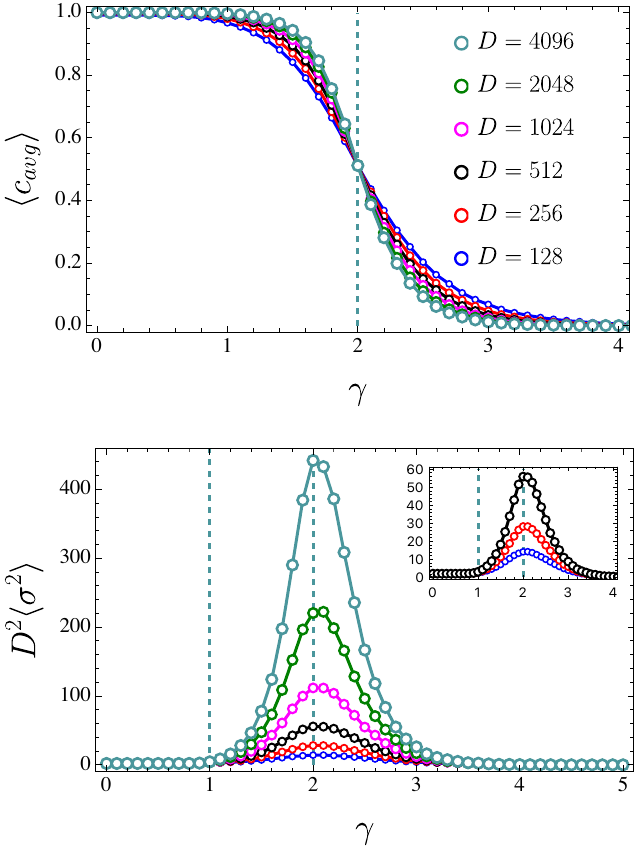}
\caption{(Top) Variation of the average of relaxation fluctuation of GUE- RP model. The crossing is clearly visible at $\gamma= 2$,  suggesting a transition at that point. (Bottom) Variation of the re-scaled ( with size) variance of relaxation fluctuation. The re-scaled variance shows a clear distinction between the chaotic and the fractal regime. Similar to the GOE case, we can identify the localized regime as the one from where the value of the re-scaled variance starts to drop.  The inset shows the zoomed-in view of the first three curves. } \label{fig:rf_all_rpgue}
\end{figure} 

As a further study, we also consider the GUE- RP model, which is given by Eq. \eqref{eqn:RPmodel} except now the matrix $B$ is a GUE matrix, whose elements, both real and imaginary parts, are drawn from a normal distribution with zero mean (for both diagonal and off-diagonal elements ) and variance is 1 and 1/2 for diagonal and off-diagonal elements respectively. It is also to be noted that for such a matrix the diagonal elements are real and off-diagonal elements are complex. The model exhibits a similar ergodic to localized transition with the fractal regime in between with corresponding transitions at $\gamma =1$ \,\text{and}\, 2 respectively.

\begin{figure}[htbp]
\includegraphics[width=0.85\linewidth]{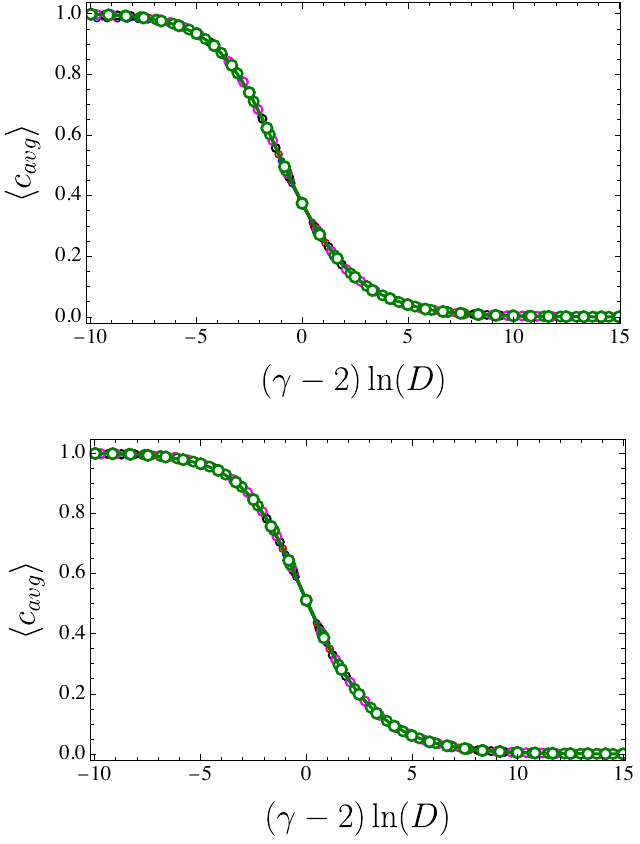}
\caption{(Top) Scaling collapse of $\braket{\cavg}$ when plotted against $(\gamma -2) \ln(D)$ for GOE- RP model. (Bottom) Scaling collapse of $\braket{\cavg}$ when plotted against $(\gamma -2) \ln(D)$ for GUE- RP model.} \label{fig:rf_collapse}
\end{figure}

We present the results of $\braket{\cavg}$ with $\gamma$ in Fig. \eqref{fig:rf_all_rpgue}. Similar to the GOE-RP model we observe the crossing of the curves at $\gamma= 2$. The behaviour of $\braket{\cvar}$ also follows the same behaviour as its GOE counterpart and is able to distinguish between the three phases. 

Finally, we revert back to the variation of $\braket{\cavg}$ with $D$. Since we have an intersection of curves at $\gamma= 2$, we can perform a data collapse. Similar to the $\braket{r}$ value \cite{PhysRevB.109.024205,pino2019ergodic}, we find that for $\braket{\cavg}$ too has the scaling collapse that can be obtained by plotting it against $(\gamma -2) \ln(D)$ for both GOE and GUE-RP model. We show this in Fig. \ref{fig:rf_collapse}. All these results exploring the correlations of eigenvectors in this system give an alternative way to study the chaotic-integrable transition along with an alternative way to explore the properties of the fractal states as well. The possibility to use $\braket{\cavg}$ and $\sigma^{2}$  to identify such transition was already pointed out in \cite{lakshminarayan1994relaxation}, based on the study of Standard map.

\section{Conclusion and Outlook} \label{sec:conclusion}
We have shown and explored in detail the relaxation fluctuation of the correlation function and their properties as outlined in detail in \cite{lakshminarayan1998relaxation,PhysRevE.56.2540} to explore chaotic properties of various systems such as mixed-field Ising model and also two types of RP model. We find that although for mixed field Ising model the eigenvalue statistics can be shown to follow RMT behaviour \cite{PhysRevLett.106.050405,PhysRevB.101.174313} the average and variance of correlation fluctuation do not follow the same and differ from the corresponding RMT behaviour. It has been already pointed out in \cite{lakshminarayan1998relaxation,PhysRevE.56.2540} and we show again in section \ref{sec:bakerrevisit} that similar behaviour can be seen for Baker's map as well where the eigenvalues follow the RMT statistics while the correlation fluctuation differs from the same. We also point out that such behaviour for both these models had been hinted at in other studies as well. Hence, we suggest that eigenvalue statistics along with the eigenvector distribution and correlations as discussed in the present framework, should be used to quantify systems as quantum chaotic. We further study a simple one-parameter model of random matrices exhibiting ergodic, fractal and localized regimes depending on its parameter, the Rosenzweig- Porter model. We study two of its variants corresponding to GOE and GUE random matrices. We observe that using both the average and the fluctuations we can identify all the three phases. While the average is just sensitive to only the fractal to ergodic transition, the variance of the fluctuations can be used to detect all three phases. Hence, using these studies we provide a novel way to study such a system which has both fractal as well as localized states and also put into firm footing the use of correlation fluctuations as a novel measure to quantify quantum chaos.

It is however to be noted that all the studies reported here are far from complete and general. All the results and the behaviour observed are with the specific choice of projection operators, which were originally chosen for their well-defined classical limits. A generalization to other operators, not necessarily projection operators is thus desirable. In these studies, we have restricted to mostly the case of $f= \frac{1}{2}$ and it would be desirable to study the variation of these quantities with partition $f$. In the case of integrable cases, it is known that the coefficients in the scaling of $\sigma_{c}$ are \emph{non-universal} and show dependence on $f$ \cite{PhysRevE.56.2540}, it would be important to study this dependence further. As the quantity $\cavg$ has been argued to be a generalization of the IPR, it is thus of much interest to further explore its features in systems exhibiting localisation. We also believe that the methods described here can be used to study in detail and quantify chaos in various other systems of present interest such as the Sachdev-Ye-Kitaev model ( see \cite{Rosenhaus:2018dtp,RevModPhys.94.035004} for a comprehensive review) and their sparse variants \cite{Xu:2020shn, Tezuka:2022mrr,  Anegawa:2023vxq, Caceres:2021nsa, Caceres:2023yoj}.

\emph{Acknowledgements:}
The author would like to thank Masaki Tezuka for the critical reading of the draft and useful suggestions to shape the paper well. The author would also like to thanks both the anonymous referees whose valuable comments improved the paper and made it more self contained. Numerical computations were performed using the computational facilities of the Yukawa Institute for Theoretical Physics. The Yukawa Research Fellowship of the author is supported by the Yukawa Memorial Foundation and JST CREST (Grant No. JPMJCR19T2). 

\appendix 
\section{Derivation for GUE}\label{appned:guededuce}
\begin{figure}[t]
\includegraphics[width=0.85\linewidth]{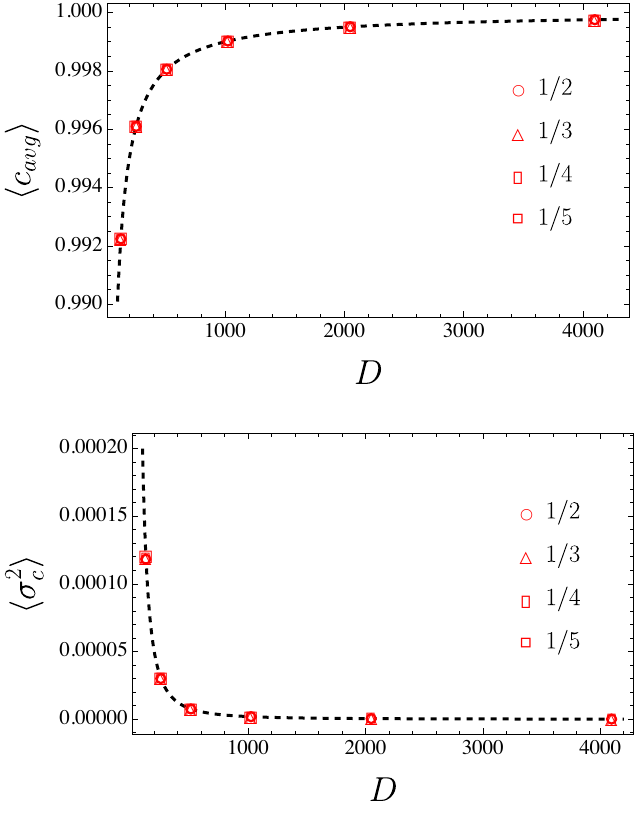}
\caption{(Top) Variation of $\braket{\cavg}$ with $D$. The dashed black line is the deduced analytic result and the red markers are the numerical results by taking different values of $f$. (Bottom) Variation of $\braket{\cvar}$ with $D$. Similar to the previous one, the dashed black line is the deduced analytic result and the red markers are the numerical results by taking different values of $f$. For both the plots we observe that the markers are on top of each other which implies the independence of the results w.r.t partition $f$, similar to the case of the GOE.} \label{fig:rf_allf}
\end{figure}

In this appendix, we provide some numerical and analytical results results for the GUE case. In Fig.\eqref{fig:rf_allf} we provide the results for $\cavg$ and $\cvar$ for various values of $f$ with system sizes $D= 128 (2048), 256 (1024), 512 (512), 1024(128), 2048(64)$ $, 4096(32)$, where the numbers in the bracket denote the number of samples used for averaging. It is important to note that similar to the case of GOE we would expect our results to be \emph{universal} in the sense that the results are independent of the partition $f$.  

Using results as shown in Fig. \eqref{fig:rf_allf}, a non-linear fitting and taking motivation from similar results for the GOE case, we can deduce the following 
\begin{align}
    \braket{\cavg}_{\text{GUE}} &= \frac{D}{D+1} \sim 1- \frac{1}{D}, \label{eqn:avgapp}\\
    \braket{\sigma_{c}^{2}}_{\text{GUE}} &= \frac{2}{D^{2}} + \mathcal{O}\left(\frac{1}{D^{4}}\right), \label{eqn:varapp}
\end{align}
where for $\cvar$ the result has been written in the same spirit as the GOE result.
It is possible to derive Eq. \eqref{eqn:avgapp} using the results for Inverse participation ratio (IPR). For an eigenstate $\ket{\alpha}$ of Hamiltonian $\mathcal{H}$ we can write the IPR for a given state $\ket{\psi}$ as 
\begin{equation}
    \mathcal{I}= D \sum_{\alpha=1}^{D} |\braket{\alpha|\psi}|^{4}
\end{equation}
In \cite{Kaplan_2007} it is stated that for the case when $\mathcal{H}$ has time-reversal symmetry, then the eigenstates $\ket{\alpha}$ are real and they are complex otherwise. Taking them to be uniformly distributed, and averaging over all the states, the results for IPR stated is 
\begin{equation}\label{eqn:iprboth}
    \frac{\mathcal{I }}{D^{2}}= \begin{cases} 
      \frac{3}{D(D+2)}, \quad  \text{for $\psi$ real}\\
     \frac{2}{D(D+1)}, \quad  \text{for $\psi$ complex}
   \end{cases}
\end{equation}
The above result then implies that $\braket{|\braket{\alpha|\psi}|^{4}}= \frac{2}{D(D+1)}$ for $\psi$ complex, where $\braket{\cdot}$ denotes ensemble average. 
In \cite{Bies_2001,Kaplan_2007} the results have been stated to be obvious. Just for completeness, we try to re-derive these results and further use them to obtain Eq. \eqref{eqn:avgapp}.\\
First, we use the hint given in \cite{Kaplan_2007} that to derive the result one can use the fact that $\braket{\alpha|\psi}$ is a complex or real number. For simplicity, we first consider the real case. What we are interested in is to find the value for $\braket{|\braket{\alpha|\psi}|^{4}}$, and we also have a normalization condition $\sum_{\alpha} |\braket{\alpha|\psi}|^{2} =1$. For the purpose of derivation, we rephrase the question as follows 
\begin{quotation}
   \emph{ If the collection of points $\{z_{i}\} \in R^{D}$ are uniformly distributed on the surface of the sphere, then what is the resulting distribution?}
\end{quotation}
The above question can alternatively be interpreted as : 
\begin{quotation}
   \emph{What is the expectation value of the sum of $4^{th}$ power of the components of a real vector which is constrained to lie on a unit Hypersphere?}
\end{quotation}
First, we note that the uniform distribution on a unit sphere of $R^{d}$ is the distribution of 
$
    \frac{z_{1}}{||z||} 
$ where $z$ are Gaussian random variables $\sim \mathcal{N}(0,\sigma)$. To this end we note, this means that $z^{2} \equiv Z$ follows Gamma distribution $\sim \Gamma(\frac{1}{2},2 \sigma)$. Thus for the case of real $z_{1}$, we want to know the distribution of 
\begin{equation}\label{eqn:unifromdist}
    \frac{z_{1}^{2}}{||z||^{2}}= \frac{Z_{1}}{Z_{1}+ \cdots + Z_{D}} 
\end{equation}
The above distribution is the ratio of two $\Gamma$ distributed variables. We now use the following properties of $\Gamma$ distribution
\begin{itemize}
    \item If $P \sim \Gamma(\alpha,\theta), Q\sim \Gamma(\beta,\theta) \implies  P+Q \sim \Gamma(\alpha +\beta,\theta)$.
    \item The distribution of $\frac{A}{A+B}$, where $A \sim \Gamma(\alpha,\theta)$ and $B \sim \Gamma(\beta,\theta)$, $A$ and $B$ are independent, is a Beta distribution with parameter $\alpha$ and $\beta$.
\end{itemize}
Using these properties, the distribution of Eq. \eqref{eqn:unifromdist} thus is Beta distribution $f_{Z}(\frac{1}{2}, \frac{n-1}{2})$.
The required expectation value $\mathbb{E}{(z_{1}^{4})}$ is then the second moment, $\mathbb{E}(Z_{1}^{2}) = \mathbb{E}(z_{1}^{4})$, of $f_{Z}(\frac{1}{2}, \frac{n-1}{2})$. The PDF of the distribution is given as
\begin{equation}
    f_{Z}(\alpha,\beta)= \frac{Z^{\alpha-1} (1-Z)^{\beta-1}}{B(\alpha,\beta)} \quad \text{for}, 0 \leq Z \leq 1.
\end{equation}
For the case of real numbers we have $\alpha= \frac{1}{2}, \beta= \frac{(D-1)}{2}$. We have to evaluate $\mathbb{E}(Z^{2})$ for this distribution to obtain our result. In fact the moments $\mathbb{E}(Z^{k})$ are easy to derive and are given as 
\begin{equation}
   \mathbb{E}(Z^{k})= \frac{B(\alpha+k,\beta)}{B(\alpha,\beta)}
\end{equation}
For real cases we thus get 
\begin{equation}
\mathbb{E}(Z^{2}) \equiv \mathbb{E}(z^{4})= \frac{3}{D(D+2)}
\end{equation}
For complex numbers, since each $z_{k}= x_{k} + i \,y_{k}$ has two degrees of freedom and there are double the total real numbers, so we want to know the distribution of 
\begin{equation}\label{eqn:unifromdistc}
    \frac{|z_{1}|^{2}}{||z||^{2}}= \frac{X_{1}+ Y_{1}}{X_{1}+ Y_{1} \cdots + X_{D}+ Y_{D} }
\end{equation}
where $|x_{i}|^{2} \equiv X_{i}, $and $|y_{i}|^{2}= Y_{i}$.
Hence for this case the Beta distribution parameters become, $\alpha=1, \beta= (D-1)$. The required moment then becomes 
\begin{equation} \label{eqn:momz4}
    \mathbb{E}(|Z|^{2})= \mathbb{E}(z^{4})= \frac{2}{D(D+1)}
\end{equation}.
We also notice that 
\begin{equation}
    z_{1}^{2}+z_{2}^{2} + \cdots + z_{D}^{2}= 1
\end{equation}
Squaring both sides and making use of the property that all the values are i.i.d we can write 
\begin{align} \label{eqn:momz2z2}
    D \braket{|z_{1}|^{4}} + D(D-1) \braket{|z_{1}|^{2} |z_{2}|^{2}} =1 \nonumber \\
    \braket{|z_{1}|^{2} |z_{2}|^{2}} = \frac{1}{D(D+1)}
\end{align}
when $z_{1} \neq z_{2}$ (notice the case when they both are equal is the evaluation of moment $\mathbb{E}(z^{4})$). We also notice that by doing the same for the real case we will obtain $\braket{z_{1}^{2} z_{2}^{2}} = \frac{1}{D(D+2)}$, which can be used for deriving the GOE result given by Eq. \eqref{eq:cavggoeresult} as well.  \\
Combining the results given by Eq. \eqref{eqn:momz4} and \eqref{eqn:momz2z2} we can write
\begin{equation}
   \braket{|z_{i}|^{2}|z_{j}|^{2}} = \frac{1 +\delta_{ij}}{n(n+1)}
\end{equation}
We can identify $z_{i} \sim \braket{\alpha_{i} \psi}$ and $z_{j} \sim \braket{\alpha_{j} \psi}$ which then give 
\begin{equation}
  \braket{|\braket{\alpha_{i}| \psi}|^{2} |\braket{\alpha_{j} |\psi}|^{2}} = \frac{1 +\delta_{ij}}{D(D+1)}
\end{equation}
where $\delta_{ij}$ factor takes care of orthogonality of $\ket{\alpha_{i}}$ and $\ket{\alpha_{j}}$ in this case. This result is also stated in \cite{Smith_2009} as well. Substituting above in Eq. \eqref{eqn:cavgfinal} and doing the summation we obtain
\begin{align}
    \braket{\cavg}_{\text{GUE}} &= \frac{D}{D+1} \sim 1- \frac{1}{D}
\end{align}
which gives us the required result, Eq. \eqref{eqn:avgapp}.
We have plotted the numerical results for both $ \braket{\cavg}_{\text{GUE}}$ and $ \braket{\cvar}_{\text{GUE}}$ in Fig.  \eqref{fig:rf_allf}, along with the derived and the deduced result. The overlapping of all the data points for different values of $f$ and the same value of $D$ indicates the independence of the results with $f$. 

\section{More results on Ising model}\label{appendix:ising}
We now consider another set of parameters called the Kim-Huse parameters \cite{PhysRevLett.111.127205}. For the mixed field Ising Hamiltonian given by Eq \eqref{eqn:hamising}
\begin{align}
H = - \sum_{i=1}^{N-1} \sigma_{i}^{z} \sigma_{i+1}^{z} - \sum_{i=1}^{N} ( h_{x}\sigma_{i}^{x}+ h_{z} \sigma_{i}^{z}),
\end{align}
The Kim-Huse parameters correspond to the maximally chaotic case, based on studies of the entanglement properties of this model
\begin{align}
    (h_x, h_z) = \left(\frac{\sqrt{5}+5}{8},\frac{\sqrt{5}+1}{4}\right) \approx (0.905, 0.809)
\end{align}
\begin{figure}[t]
\includegraphics[width=0.95\linewidth]{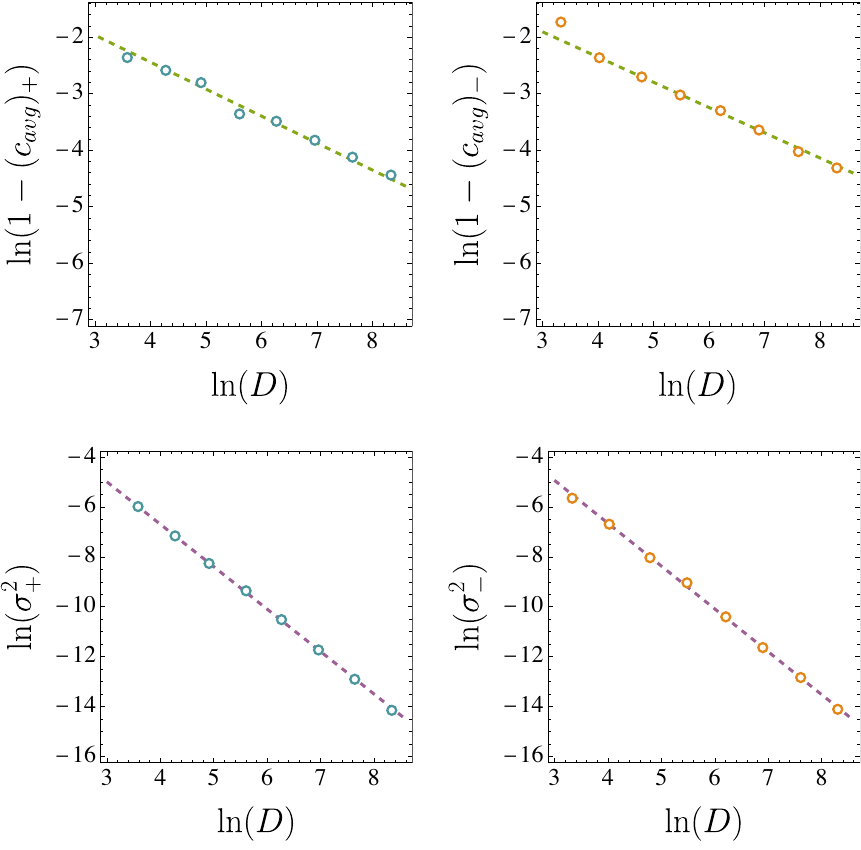}
\caption{(Top) The logarithmic deviation of average fluctuation from unity for the parity $+1$ and $-1$ sector respectively, for the mixed field Ising model with Kim-Huse parameters. (Bottom) Variation of the logarithmic variance of the relaxation for the mixed field Ising model, with Kim-Huse parameters, for parity $+1$ and $-1$ sector respectively. In both the plots the dashed line is the best-fit line.} \label{fig:rf_isingkh}
\end{figure}
For the chaotic regime, the results for the deviation of average relaxation from unity are shown in Fig (Top). \eqref{fig:rf_isingkh} for both the parity sectors. We observe the following power-law-like behaviour of average fluctuation with $D$ 
\begin{equation}
(\cavg)_{\pm} = 1- \frac{\alpha_{\pm}}{ D ^{\beta_{\pm}}}, 
\end{equation}
where $\pm$ denotes the $+1$ and $-1$ parity sectors respectively with $(\alpha_{+},\beta_{+)} \approx (\frac{3}{5},\frac{1}{2}),(\alpha_{+},\beta_{+)} \approx (\frac{3}{5},\frac{2}{5})$, which deviates from the RMT predictions more than the BCH parameters used in section \ref{sec:Ising}.
\bibliography{references}

\end{document}